\documentclass[10pt,twocolumn,amsmath,amssymb,prb]{revtex4}
\usepackage[dvips]{graphicx}
\usepackage{amsmath}

\newcommand{\ie}{i.e.\ }

\newcommand{\wn}{wall-normal }

\begin{document}

\title{Linear feedback control of transient energy growth\\ and control performance limitations\\ in subcritical plane Poiseuille flow}
\author{F. Martinelli\footnote{Present address: LadHyX, \'Ecole Polytechnique, F-91128 Palaiseau, France. e-mail: fulvio.martinelli@polimi.it}}
\affiliation{Politecnico di Milano, Dipartimento di Ingegneria Aerospaziale,\\
Via La Masa 34, Milano, Italy}
\author{M. Quadrio}
\affiliation{Politecnico di Milano, Dipartimento di Ingegneria Aerospaziale,\\
Via La Masa 34, Milano, Italy}
\author{J. McKernan}
\affiliation{Experimental \& Computational Laboratory for the Analysis of Turbulence, King's College London, \\
Strand, London WC2R 2LS, U.K.}
\author{J.F. Whidborne}
\affiliation{Department of Aerospace Sciences, Cranfield University, \\
Cranfield, Bedfordshire, MK43 0AL, U.K.}

\begin{abstract}
Suppression of the transient energy growth in subcritical plane Poiseuille flow via feedback control is addressed.
It is assumed that the time derivative of any of the velocity components can be imposed at the walls as control input, and that full-state
information is available. We show that it is impossible to design a linear state-feedback controller that leads to a closed-loop flow system 
without transient energy growth. In a subsequent step, full-state feedback controllers -- directly targeting the transient growth mechanism -- are designed,
using a procedure based on a Linear Matrix Inequalities approach. 
The performance of such controllers is analyzed first in the linear case, where comparison to previously 
proposed linear-quadratic optimal controllers is made; further,
transition thresholds are evaluated via Direct Numerical Simulations of the controlled 
three-dimensional Poiseuille flow against different initial conditions of physical interest,
employing different velocity components as wall actuation. 
The present controllers are effective in increasing the transition thresholds in closed loop, with varying degree of performance depending
on the initial condition and the actuation component employed.
\end{abstract}
\maketitle

\section{Introduction}
Transient energy growth has been recognized as a possible mechanism explaining subcritical transition in wall-bounded flows;
in fact, subcritical flows may experience large transient amplifications of the energy of perturbations, that could trigger
nonlinear mechanisms and eventually lead to transition to turbulence\cite{butler-farrel-1992, reddy-henningson-1993, schmid-henningson-1994}.

In viscous shear flows, transient energy growth is related to the 
non-normality of the linearized Navier-Stokes operator with respect to the energy inner product\cite{schmid-henningson-2001, schmid-2007}.
In the last few years, several investigators attempted to reduce the transient growth phenomenon in Poiseuille and boundary 
layer flows by employing wall actuation and applying linear control theory to an appropriate discretization of the linearized equations. 
In their seminal work on feedback control of instabilities in two-dimensional Poiseuille flow, Joshi et al.\cite{joshi-speyer-kim-1997} 
employed a compensator in the form of a constant-gain integral feedback, and demonstrated stabilization of the linearly
unstable flow as well as attenuation of finite amplitude disturbances. They further pointed out that transient amplifications
in the flow energy may not be properly detected by the sensors, and also that the control itself may trigger nonlinear
mechanisms by introducing transient disturbances on short times. 
Leveraging a state-space formulation obtained after discretization of the boundary-controlled Orr-Sommerfeld-Squire equations,
optimal and robust control theory was applied to transitional channel flows by Bewley \& Liu\cite{Bewley-1998-305} for a single 
wavenumber pair and by H\"ogberg et al.\cite{Hogberg-2003-149} for a large array of wavenumber pairs,
leading to a reduction of the maximum transient energy growth as well as an increase in transition thresholds. 
It has been recently shown\cite{kim-lim-2000} that the linear coupling term in the Orr-Sommerfeld-Squire equations 
plays a role not only in the non-normal behavior of the small perturbation dynamics, but also in the self-sustaining
process of near-wall, low Reynolds number turbulence. This evidence led investigators to test in turbulent channel flows the optimal
controllers designed on linearized flow models, and encouraging results have been obtained in terms 
of drag reduction\cite{lee-etal-2001, bewley-moin-temam-2001, kim-2003, bewley-2001}.
Feedback control of non-modal disturbances in boundary layer flows has been recently considered by Corbett \& Bottaro\cite{corbett-bottaro-2001} 
in the framework of optimal control theory, while Zuccher et al.\cite{zuccher-luchini-bottaro-2004} applied 
steady suction in the attenuation of the growth of given optimal disturbances in a Blasius boundary layer.

Although it has been demonstrated\cite{bewley-2001} that optimal and robust control laws are well suited
for reducing the non-normal behavior of fluid flow systems, to date no feedback control law has been devised
with the capability of ensuring closed-loop monotonic stability, when boundary actuation is employed. 
It is therefore natural to ask whether such performance can be obtained with a linear feedback, and further 
which control techniques are available to directly target the transient growth mechanism.
In the controls literature, the transient amplification of certain norms of the state bears the name of \emph{peaking phenomenon}
and the monotonic stability requirement is generally referred to as \emph{strict dissipativity}. 
Active controllers with the capability of targeting transient norm amplifications have received attention
in the analysis of a class of partially linear cascade systems\cite{sussmann-kokotovic-1991}, and more recently
in conjunction with a Linear Matrix Inequality (LMI) approach\cite{plischke-wirth-2004, plischke-2005}.
In a very recent paper, Whidborne \& McKernan\cite{Whidborne-2007-1762} extended these results giving conditions on the existence of a feedback
controller ensuring the strict dissipativity of the closed-loop system. These results were then exploited 
by Whidborne et. al\cite{WhidborneMckernanPapadakisJSCE2008}, who considered the feedback control of a single wavenumber pair 
in plane Poiseuille flow via LMI design and wall-normal blowing and suction, and by Martinelli et al.\cite{martinelli-etal-2009}, 
where LMI-based feedback controllers have been designed and tested for an array of wavenumber pairs.
The present paper aims at expanding and completing these recent results, 
by showing first that it is impossible to design a linear state-feedback controller
ensuring the plane Poiseuille flow -- controlled via wall transpiration with any velocity component -- to be strictly dissipative. 
In a second step, feedback control laws are designed using an LMI technique, for an array of wavenumber pairs; their performance
is compared against that of optimal controllers in the linear case, and furthermore
closed-loop transition thresholds are evaluated for optimal initial conditions in 
the form of a pair of oblique waves and antisymmetric streamwise vortices, at the Reynolds number $Re=2000$, using different
velocity components as wall actuators.

\section{Model of the system}

We consider the dynamics of three-dimensional small perturbations to the laminar Poiseuille solution in a plane channel. 
A Cartesian coordinate system is introduced, where $x$, $y$ and $z$ denote the streamwise, wall-normal and spanwise directions, and 
$u$, $v$, $w$ denote the corresponding perturbation velocity components.
The Navier-Stokes equations, linearized about the laminar solution $U(y) = U_p(1-(y/\delta)^2)$,
are non-dimensionalized with the centerline velocity $U_p$ and the channel half-width $\delta$,
and rewritten in the form of a single equation for $v$ one-way coupled to an equation for the wall-normal vorticity 
$\eta = {\partial u}/{\partial z} - {\partial w}/{\partial x}$.
Fourier transformation in $x$ and $z$ direction yields the well known Orr-Sommerfeld-Squire form:
\begin{equation}
\begin{split}
\Delta \dot{\tilde{v}} &= [-j \alpha U \Delta + j\alpha U'' + \Delta\Delta/Re] \tilde{v} \\
\dot{\tilde{\eta}} &= [-j\beta U'] \tilde{v} + [-j\alpha U + \Delta/Re] \tilde{\eta}
\end{split}
\label{eq:oss}
\end{equation}
at the wavenumber pair $(\alpha,\beta)$. 
Here, the tilde denotes Fourier coefficients, the dot denotes time derivative, the prime denotes $y$ differentiation, $\kappa^2 = \alpha^2 + \beta^2$, 
$j$ is $\sqrt{-1}$, and $\Delta = d^2/dy^2 - \kappa^2$.

We select boundary conditions representing time-varying wall transpiration on any of the velocity components
at the two channel walls (``vectorized transpiration''). In turn, this results in inhomogeneous Dirichlet and 
Neumann conditions on $\tilde{v}$, as well as inhomogeneous Dirichlet conditions on $\tilde{\eta}$: 
\begin{equation}
\begin{aligned}
        \tilde{v}(y=\pm1,t)  & =  \tilde{v}_{u,l}(t), \\
        \partial \tilde{v}/\partial y(y=\pm{1},t) & = \tilde{v}_{u,l}'(t), \\
        \tilde{\eta}(y=\pm{1},t) & = \tilde{\eta}_{u,l}(t). 
\end{aligned}
\end{equation}
A standard lifting procedure\cite{Hogberg-2003-149} is then employed, i.e. 
the unknowns are rewritten as homogeneous components (satisfying homogeneous boundary conditions)
plus inhomogeneous components thus:
\begin{equation}
\begin{split}
&\tilde{v}(y,t) = \tilde{v}_h(y,t) + f_u(y)\tilde{v}_u(t) + f_l(y)\tilde{v}_l(t) + \ldots \\
&\qquad\quad\ldots + g_u(y)\tilde{v}_u'(t) + g_l(y)\tilde{v}_l'(t) \\ 
&\tilde{\eta}(y,t) = \tilde{\eta}_h(y,t) + h_u(y)\tilde{\eta}_u(t) + h_l(y)\tilde{\eta}_l(t) ,
\end{split}
\label{eq:cvar}
\end{equation}
where $f_u, f_l, g_u, g_l, h_u, h_l$ are polynomials in $y$ chosen to satisfy unitary boundary conditions for each lifted
component appropriately.

We discretise the homogeneous components $\tilde{v}_h$ and $\tilde{\eta}_h$
in \wn direction using a modified Chebyshev series cardinal function basis
\begin{equation}
\begin{aligned}
\tilde{v}_h(y,t) &= \sum_{n=0}^{N-4} \Gamma_n^{DN}(y) a_{v,n}(t) \\
\tilde{\eta}_h(y,t) &= \sum_{n=0}^{N-2} \Gamma_n^D(y) a_{\eta,n}(t)
\end{aligned} 
\label{eq:cheb_basis}
\end{equation}  
where the modified Chebyshev functions $\Gamma_n^{DN}(y)$ and $\Gamma_n^{D}(y)$
implicitly enforce the required 
homogeneous Dirichlet and Neumann boundary conditions, lead to good conditioning of the discrete Laplacian operator, and no
spurious modes are generated \cite{thesis}.
The Orr-Sommerfeld-Squire equations are then evaluated on a set of Gauss-Lobatto collocation points in $y$ direction
and rearranged to have the time rate of change of actuation velocity as an input\cite{Hogberg-2003-149}. This results in the
the linear time-invariant system model\cite{Skogestad01} 
\begin{equation*}
    \dot{\overline{\textbf{x}}}(t) = \overline{\textbf{A}}\overline{\textbf{x}}(t) + \overline{\textbf{B}}\textbf{u}(t),\quad  
					\overline{\textbf{x}}(0)=\overline{\textbf{x}}_0,
\end{equation*}
where $\overline{\textbf{A}}$, $\overline{\textbf{B}}$ are constant system and input matrices, and
$\textbf{u}$, $\overline{\textbf{x}}$ are respective input and state vectors. As noted in previous work\cite{Hogberg-2003-149},
the natural outcome of this procedure is an augmented state-space form, where additional integrators 
associated to the values of the input velocity components at the walls are explicitly introduced. Consequently,
the \emph{open-loop} dynamics of the system above (i.e. setting $\overline{B}=0$) is different from the dynamics of the
original, \emph{unactuated} system (where the velocity at the walls is fixed by the no-slip condition). 
This terminology will be used throughout the paper to distinguish between the two cases.

The kinetic energy per unit mass of flow perturbations in the volume $V$  
\begin{equation*}
E = \frac{1}{2V} \int_V u^2 + v^2 + w^2 \, dV
\end{equation*}
can be expressed as a function of the state vector $\overline{\textbf{x}}$ using the continuity equation, the definition of $\eta$, and
Parseval's identity:
\begin{equation*}
E = \sum_{(\alpha,\beta)} \tilde{E}(\alpha,\beta) = \sum_{(\alpha,\beta)} \overline{\mathbf{x}}^H \mathbf{Q}(\alpha,\beta) \overline{\mathbf{x}} 
\end{equation*} 
where the matrix $\mathbf{Q}$ is an Hermitian, positive definite matrix; here, the superscript $H$ denotes conjugate transpose.
In the following, we have transformed the state vector via the change of variable $\textbf{x}=\textbf{C}\overline{\textbf{x}}$
($\textbf{C}$ being the Cholesky factor of $\textbf{Q}$)
in order to rewrite the system dynamics as 
\begin{equation}
    \dot{\textbf{x}}(t) = \textbf{A}\textbf{x}(t) + \textbf{B}\textbf{u}(t),\quad  
					\textbf{x}(0)=\textbf{x}_0,
\label{eq:ltiABC}
\end{equation}
such that the system energy is directly given by the Euclidean norm $\textbf{x}^H\textbf{x}$\cite{Skogestad01, schmid-2007}.

\section{Closed-loop monotonic stability}
\label{sec:monotonic}

We consider the linear time-invariant system model (\ref{eq:ltiABC})
and further assume that $\textbf{B}^{H}\textbf{B}>0$, that is $\textbf{B}$ has full
column rank (i.e. all the actuators are independent -- a condition which is trivially satisfied in the present problem). 
Contraction analysis for this kind of system has been presented by Whidborne \& McKernan\cite{Whidborne-2007-1762} 
and, in the general case of nonlinear systems, by Lohmiller \& Slotine\cite{lohmiller-slotine-1998}.
In particular, referring to the linear case, it has been shown\cite{Whidborne-2007-1762} that there exists a static state-feedback controller 
$\textbf{u}=\textbf{K}\textbf{x}$, where $\textbf{K}$ is a constant
matrix, such that the closed-loop system has strict dissipativity 
(\ie energy $\textbf{x}^H\textbf{x}$ decays monotonically from all initial conditions $\textbf{x}_0$), if and only if
\begin{equation}
\textbf{B}^{\bot} \left(\textbf{A}+\textbf{A}^{H}\right) \textbf{B}^{\bot H}   <0 \text{ or } \textbf{B}\textbf{B}^{H}>0,
\label{eq:criterion}
\end{equation}
where $\textbf{B}^{\bot}$ is the left null space of $\textbf{B}$.
Additionally\cite{Whidborne-2007-1762}
if no static controller that achieves
strict dissipativity exists, then no dynamic state-feedback controller, where $\textbf{u}$ is given from $\textbf{x}$ by the dynamic system
\begin{eqnarray}\label{eq:ltiK}
    \dot{\textbf{x}}_k(t) &=& \textbf{A}_k \textbf{x}_k(t) + \textbf{B}_k \textbf{x}(t),\quad  \textbf{x}_k(0)=\textbf{x}_{k0},\\
    \textbf{u}(t) &=& \textbf{C}_k \textbf{x}_k(t) + \textbf{D}_k \textbf{x}(t),
\end{eqnarray}
where $\textbf{A}_k$,$\textbf{B}_k,\textbf{C}_k$ and $\textbf{D}_k$ are constant matrices and $\textbf{x}_k$ are controller states, exists either.

It is immediate to verify that the second criterion in \eqref{eq:criterion} is never satisfied in the present system, as the Hermitian matrix 
$\textbf{B}\textbf{B}^{H}$ is never positive definite but it is always positive semidefinite, because the dimension 
of the input vector is always smaller than that of the state vector. In order to have $\textbf{B}\textbf{B}^{H}>0$,
a number of independent actuators equal to the number of flow states is required; this is a situation that
is unlikely to occur in practical flow control problems, where normally actuators are placed 
at the walls. Even when volume forcing is available, this condition is unlikely to be satisfied, 
since practical volume forces are not as flexible as to enforce an arbitrary force distribution in the entire 
flow domain at any time instant.

\begin{figure}[!htbp]
\begin{center}
\includegraphics[angle=-90, width=0.45\textwidth]{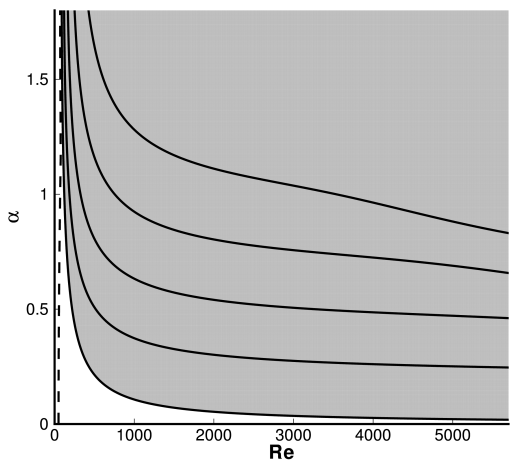}
\end{center}
\caption{Numerical verification of the first algebraic criterion in Eq. \eqref{eq:criterion}.
Lines: contours at constant $\lambda_{max}(Re,\alpha)$, at $\beta=0$. Levels are from $-0.1$ to $0.4$ by $0.1$ increments;
dashed line is negative value. The shaded area corresponds to the region where the unactuated system is not monotonically stable, i.e.
admits transient energy growth. The contour $\lambda_{max}(Re,\alpha)=0$ lies on the boundary of the region, indicating
that no state-feedback controller can be designed to ensure strict dissipativity of the closed-loop system when the
unactuated system is not strictly dissipative. Results obtained with $N=100$, $N_t=50$.}
\label{fig:alfa_Re}
\end{figure}

\begin{figure}[!htbp]
\begin{center}
\includegraphics[angle=-90, width=0.45\textwidth]{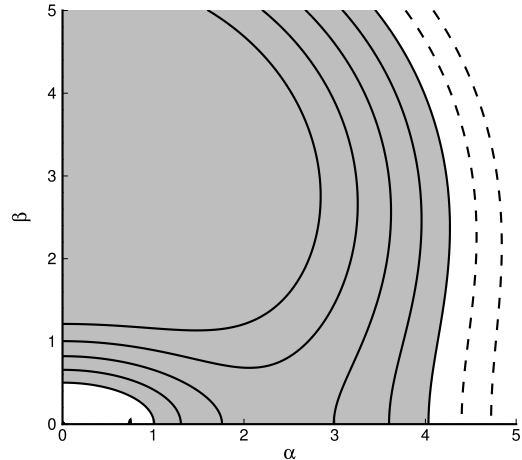}
\end{center}
\caption{Numerical verification of the first algebraic criterion in Eq. \eqref{eq:criterion}. Results for $Re=120$, levels are
	from $-0.1$ to $0.2$ by $0.05$ increments. For details see caption of Fig. \ref{fig:alfa_Re}.}
\label{fig:alfa_beta_Re}
\end{figure}

The first algebraic criterion in Eq. \eqref{eq:criterion} is equivalent to requiring that
the portion of the system dynamics which is not accessible by the controls must be dissipative. 
Verifying this criterion is not trivial. Here we evaluate it numerically, in order 
to identify those regions in the $(\alpha,\beta,Re)$ parametric space where subcritical Poiseuille flow may be rendered
monotonically stable by feedback transpiration.
To this aim, the state-space model \eqref{eq:ltiABC} is first obtained on a fine grid and, as suggested by Reddy \& 
Henningson\cite{reddy-henningson-1993}, a limited number $N_t$ of eigenfunctions is retained, discarding those corresponding to highly
damped and poorly resolved eigenvalues. 
Properly rescaling the variables such that energy is written as an Euclidean norm
leads to a reduced order model $\textbf{A}_r$,$\textbf{B}_r$, and the negative-definiteness of the corresponding
matrix $\textbf{B}_r^{\bot} \left(\textbf{A}_r+\textbf{A}_r^{H}\right) \textbf{B}_r^{\bot H}$
in \eqref{eq:criterion} is verified by computing its maximum (real) eigenvalue $\lambda_{max}$. 
Figure \ref{fig:alfa_Re} shows the present result on the $(Re,\alpha)$ plane for $\beta=0$, 
along with the well-known result on the transient growth dependence 
in plane Poiseuille flow\cite{reddy-henningson-1993} (i.e. the unactuated case). 
The white area corresponds to the domain where the unactuated system is monotonically stable,
while the shaded area is the region where the unactuated system admits transient energy growth. Solid lines
correspond to isocontours of $\lambda_{max}$, and it appears that the contour $\lambda_{max}=0$ lies on the very boundary between the 
shaded and white area, implying that the Hermitian matrix
$\textbf{B}_r^{\bot} \left(\textbf{A}_r+\textbf{A}_r^{H}\right) \textbf{B}_r^{\bot H}$ is indefinite when the unactuated system is 
not monotonically stable. 
An analogous result is reported in Fig. \ref{fig:alfa_beta_Re}, where isocontours of $\lambda_{max}$ are reported 
on the plane $(\alpha,\beta)$, at $Re=120$; again, the contour $\lambda_{max}=0$ lies on the boundary between the regions
of monotonic and non-monotonic stability.
From the aforestated theorem\cite{Whidborne-2007-1762}, this implies that it is not possible to design a state-feedback controller that ensures the
closed loop Poiseuille flow to be monotonically stable, when the corresponding unactuated flow is not.

This result shows an inherent limitation in the feedback control of the transient growth mechanism, when 
vectorized wall transpiration in terms of time rate of change of zero net-mass flux blowing/suction is employed. 
Note that vectorized transpiration, although being rather idealized, exploits all the degrees of freedom available for
boundary control in the present problem; therefore, the present result is representative of a limiting situation.
In a practical setting, actuator-dependent constraints may introduce
additional mechanisms restricting the control authority even more.
It is also worth mentioning that strict dissipativity is a rather tough requirement for a controller bound to
operate on a largely underactuated flow, and in fact an approach based on strict dissipativity is in general
quite conservative, i.e. some energy growth is tolerable in transition control. 
Finally, it should be emphasized that the present analysis is limited to linear feedback laws, and that performance
of nonlinear controllers may be more promising. For example, it has been shown\cite{hogberg-bewley-henningson-2003} that 
introducing a nonlinearity in the form of gain-scheduling on full-state feedback laws led to relaminarization of low-$Re$
turbulence even employing wall actuation only, and that adjoint-based optimization on the nonlinear turbulent flow 
can be successfully employed in feedback relaminarization \cite{bewley-moin-temam-2001}.

\section{Upper-bound minimizing feedback controller}

In order to design a state-feedback controller with the capability of targeting the transient growth mechanism directly,
an estimate of the maximum transient growth is required. Such estimate is obtained as an upper bound on the maximum 
growth via Lyapunov theory.
For the linear, time invariant, asymptotically stable system:
\begin{equation*}
\mathbf{\dot{x}} = \mathbf{A} \mathbf{x}, \qquad \mathbf{x}(0) = \mathbf{x_0}, 
\end{equation*}
it can be shown that an upper bound on the maximum transient growth $G$ is given by \cite{Whidborne-2007-1762,WhidborneMckernanPapadakisJSCE2008} :
\begin{equation*}
G_u = \lambda_{max}(\mathbf{P})\lambda_{max}(\mathbf{P}^{-1}) \ge G,
\end{equation*}
where $\mathbf{P}=\mathbf{P}^H > 0$ satysfies the Lyapunov inequality
\begin{equation*}
\mathbf{PA} + \mathbf{A}^H \mathbf{P} < 0.
\end{equation*}
A minimal upper bound can be obtained by solving the following minimization problem \cite{boyd-etal-1994} :
\begin{equation}
\begin{aligned}
& \min \gamma : \\
& \mathbf{PA} + \mathbf{A}^H \mathbf{P} < 0, \quad \mathbf{P}=\mathbf{P}^H>0 \\
& \mathbf{I} < \mathbf{P} < \gamma \mathbf{I},
\end{aligned}
\label{eq:lmiopen}
\end{equation}
where the last inequality ensures $\gamma > G_u$.
The problem stated in Eq. \eqref{eq:lmiopen} is a LMI generalized eigenvalue problem, and standard solution methods 
based on interior point algorithms are available\cite{nesterov-nemirovskii-1994}.

An analogous problem to that stated in Eq. \eqref{eq:lmiopen} can be obtained if the feedback minimization of the
upper bound is of interest. 
Indeed, let us consider the system \eqref{eq:ltiABC} along with a state-feedback control 
law in the form $\mathbf{u}=\mathbf{Kx}$, so that in closed-loop the system dynamics is described by:
\begin{equation*}
\mathbf{\dot{x}} = (\mathbf{A} + \mathbf{BK}) \mathbf{x}, \qquad \mathbf{x}(0) = \mathbf{x_0}. 
\end{equation*}
Leveraging the additional degrees of freedom due to the controller gains $\mathbf{K}$, we move to minimizing the
closed-loop upper bound on the maximum transient growth.
The associated Lyapunov inequality now reads:
\begin{equation*}
\mathbf{PA} + \mathbf{A}^H\mathbf{P} + \mathbf{PBK} + \mathbf{K}^H\mathbf{B}^H\mathbf{P}^H < 0;
\end{equation*}
this inequality can be rewritten in the LMI form by recalling that
a similarity transformation preserves the eigenvalues. Therefore, defining $\mathbf{Q} = \mathbf{P}^{-1}$ and $\mathbf{Y} = \mathbf{KQ}$,
the closed-loop upper-bound minimization problem can be written as \cite{boyd-etal-1994} 
\begin{equation}
\begin{aligned}
& \min \gamma : \\
& \mathbf{AQ} + \mathbf{QA}^H + \mathbf{BY} + \mathbf{Y}^H \mathbf{B}^H < 0, \quad \mathbf{Q}=\mathbf{Q}^H>0 \\
& \mathbf{I} < \mathbf{Q} < \gamma \mathbf{I}, \\
& \begin{pmatrix}
	\mathbf{Q} & \mathbf{Y}^H \\
	\mathbf{Y} & \mu^2 \mathbf{I}
  \end{pmatrix} > 0,
\end{aligned}
\label{eq:lmiclosed}
\end{equation}
where the last, additional inequality ensures a limit in the control effort in the form $\max_{t\ge0} ||\mathbf{u}||^2 < \mu^2$. 
The problem \eqref{eq:lmiclosed} has to be solved for $\mathbf{Q}$, $\mathbf{Y}$ and $\gamma$; 
controller gains are obtained from $\mathbf{K} = \mathbf{YQ}^{-1}$.
This problem is again a LMI generalized eigenvalue problem, that can be solved using standard methods \cite{nesterov-nemirovskii-1994}.

\section{Results and discussion}

LMI controllers are designed wavenumber-wise, using the model of the system \eqref{eq:ltiABC} and the
design equations \eqref{eq:lmiclosed}.
In particular, we consider each actuation component ($u$, $v$ or $w$ on both walls) independently; 
a limit on the control effort $\mu=10$, kept constant in wavenumber space, is used in the design of all controllers. 
In the design procedure, the linear equations pertaining to each wavenumber pair are discretized using $N=100$ Chebyshev polynomials, 
and modal truncation at $N_t = 54$ (the maximum affordable size of the computational problem) 
is performed prior to the actual solution of the generalized eigenvalue problem \eqref{eq:lmiclosed}.
In addition to removing poorly resolved dynamics, modal truncation proved to be necessary due to the exacting memory requirements
of the existing LMI solvers (scaling as $\approx N_t^6$); in performing modal truncation, it was thoroughly verified that the reduced order model
preserves the linear transient energy growth of the unactuated system.

\subsection{Linear analysis}

The performance of LMI controllers is evaluated first in the linear setting at $Re=2000$; 
in particular, perturbations at two representative wavenumber pairs are considered, namely, an oblique wave $(\alpha=1,\beta=1)$ and a streamwise vortex 
$(\alpha=0,\beta=2)$. In the results reported here, the effectiveness of different actuation components is also addressed.

\begin{figure}[!htbp]
\begin{center}
\includegraphics[width=0.45\textwidth]{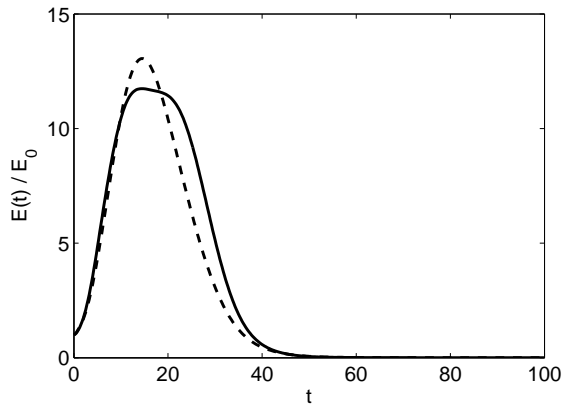}
\end{center}
\caption{Time evolution of the perturbation energy for the closed-loop system controlled using LMI gains (--) and LQR gains ($-\,-$), actuation using
$v$. The initial condition for the two systems is the respective closed-loop optimal perturbation, and the LQR gains are iteratively designed so that
the energy expense over the simulated time horizont for the two closed-loop systems is the same. Results for ($\alpha=1,\beta=1$) at $Re=2000$.}
\label{fig:lmi-lqr-closedloop-ow}
\end{figure}

\begin{figure}[!htbp]
\begin{center}
\includegraphics[width=0.45\textwidth]{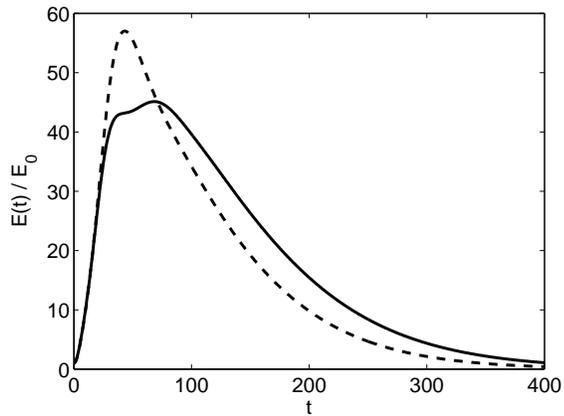}
\end{center}
\caption{Time evolution of the perturbation energy for the closed-loop system controlled using LMI gains (--) and LQR gains ($-\,-$), actuation using
$w$. The initial condition for the two systems is the respective closed-loop antisymmetric (with respect to the $y=0$ plane) optimal perturbation, and the LQR gains are iteratively designed so that
the energy expense over the simulated time horizont for the two closed-loop systems is the same. Results for ($\alpha=0,\beta=2$) at $Re=2000$.}
\label{fig:lmi-lqr-closedloop-sv}
\end{figure}

The analysis reported in sec. \ref{sec:monotonic} shows that the closed-loop system will have a non-normal behavior;
it is therefore natural to contrast the maximum closed-loop transient growth $G$ with the open-loop one, in order to verify that a consistent reduction
in $G$ is obtained via the minimization in \eqref{eq:lmiclosed}. Results, obtained for the same truncated system used in the design and for the two wavenumber pairs 
considered, are reported in table \ref{tab:contrastG}. It is shown that, in both the oblique wave case and the streamwise vortex case, solution of \eqref{eq:lmiclosed} leads to
a closed-loop system experiencing a reduced maximum transient energy growth. In particular, for the oblique wave case, $v$ is the most effective actuation component, whereas
actuating with $w$ is most effective in the streamwise vortex case.
\begin{table}
\begin{center}
\begin{tabular}{|c|c|c|c|c|}
\hline
 & \multicolumn{2}{|c|}{Oblique waves} & \multicolumn{2}{|c|}{Streamwise vortices} \\
 & Open-loop & Closed-loop & Open-loop & Closed-loop  \\
\hline
\hline
$u$ - actuation & $67.62$ & $33.04$ & $785.93$ & $635.83$ \\
\hline
$v$ - actuation & $66.81$ & $11.84$ & $33126.72$ & $129.23$\\
\hline 
$w$ - actuation & $67.13$ & $32.70$ & $163123.13$ & $85.90$ \\
\hline
\hline
\end{tabular}
\end{center}
\caption{Maximum transient energy growth $G$ for the open-loop and closed-loop cases, for the oblique wave case ($\alpha=1$,$\beta=1$) and the streamwise vortex
	case ($\alpha=0$,$\beta=2$), using different actuation components on both walls. $Re=2000$.}
\label{tab:contrastG}
\end{table}
The performance of the present controllers is further compared against full-state controllers designed with
the LQR approach\cite{Skogestad01}, considered in a similar transition problem
by Hogberg et al.\cite{Hogberg-2003-149}. The aim of the LQR control is the minimization of the time integral of the perturbation energy, while
keeping the time integral of the control effort as low as possible; in fact, the control objective is
given in terms of the closed-loop minimization of a functional in the form
\begin{equation*}
J = \int_0^{+\infty} \mathbf{\overline{x}}^H \mathbf{Q} \mathbf{\overline{x}} + \rho \mathbf{u}^H \mathbf{u} \, dt.
\end{equation*}
This is substantially different from the control objective of the present LMI formulation \eqref{eq:lmiclosed}, which considers bounds on the disturbance energy
and control expenditure; therefore -- at a fixed control expenditure -- LMI controllers can be used
to estimate a possible best performance (in terms of peaking suppression) of other control
strategies. In order to present a fair comparison between the LQR and the LMI formulation, we iteratively design
and test a LQR controller keeping $\mathbf{Q}$ fixed (the same used in LMI design) and $\rho$ as a free parameter, 
and we evaluate the integral of the control energy $\int_0^{\infty} \mathbf{u}^H \mathbf{u}\,dt$ in closed loop
until it matches the value computed for the LMI controller. 
The closed-loop systems, controlled via both LQR and LMI gains, are tested against the respective optimal perturbations, using $v$-actuation for the
oblique wave case and $w$-actuation for the streamwise vortex case (the best performance cases reported in table \ref{tab:contrastG}); further, for the 
streamwise vortex we consider the antisymmetric (with respect to the $y=0$ plane) optimal perturbation. The time evolution of the perturbation 
energy is displayed in fig. \ref{fig:lmi-lqr-closedloop-ow} and \ref{fig:lmi-lqr-closedloop-sv}. Results show that, at a given control expenditure, the worst-case initial condition
for the LMI-controlled system experiences a lower amplification than the corresponding perturbation for the LQR-controlled system; the peak for the LMI-controlled
system occurs at later times for the streamwise vortex case. Despite the mild reduction in maximum amplification, the results shown here suggest that a control design technique directly targeting the
growth mechanism is able to better exploit the degrees of freedom in the controller to achieve a minimal transient peaking of the energy.

\begin{figure}[!htbp]
\begin{center}
\includegraphics[width=0.45\textwidth]{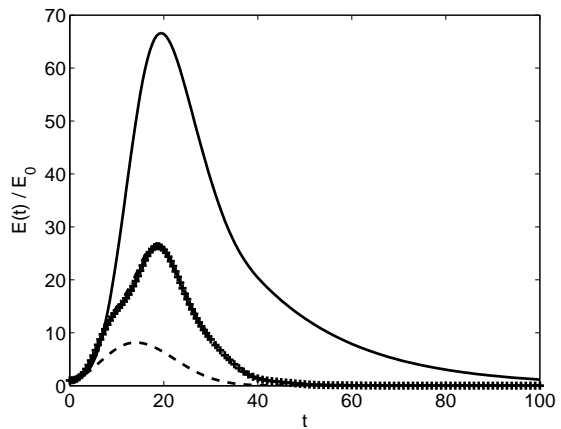}
\end{center}
\caption{Linear dynamics of the perturbation energy for the unactuated case (--), and closed-loop case actuating with $u$ ($-\,\cdot$),
$v$ ($-\,-$), $w$ ($-\,+$). Results for ($\alpha=1,\beta=1$) at $Re=2000$; in all cases, the initial condition is the optimal disturbance for the unactuated flow.}
\label{fig:lin-ow}
\end{figure}

\begin{figure}[!htbp]
\begin{center}
\includegraphics[width=0.45\textwidth]{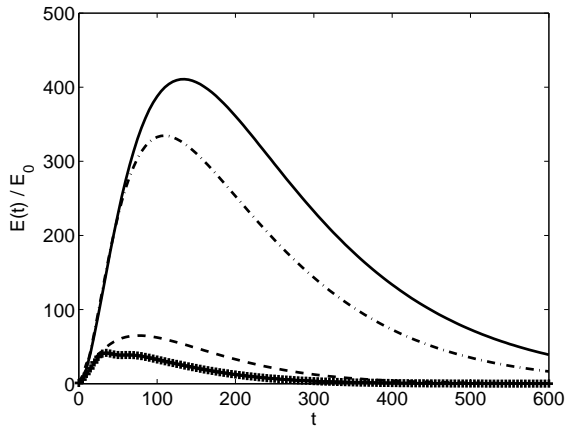}
\end{center}
\caption{Linear dynamics of the perturbation energy for the unactuated case (--), and closed-loop case actuating with $u$ ($-\,\cdot$),
$v$ ($-\,-$), $w$ ($-\,+$). Results for ($\alpha=0,\beta=2$) at $Re=2000$; in all cases, the initial condition is the antisymmetric optimal disturbance for the unactuated flow.}
\label{fig:lin-sv}
\end{figure}

The performance of LMI controllers is also evaluated against optimal initial conditions for the unactuated flow, and results are reported in
fig. \ref{fig:lin-ow} and \ref{fig:lin-sv}. 
Since the closed-loop system has additional state equations associated with the dynamics of wall velocity components used as actuators,
these values are set to zero, assuming that at initial time the open-loop flow satisfies the no-slip and no-transpiration condition at the walls.
In particular, for the oblique wave case, it is shown that actuating with $v$ reduces the maximum amplification
of the optimal disturbance by a factor $\approx8.2$, whereas a less effective reduction (by a factor $\approx2.5$)
is obtained using $u$ or $w$. Further, the growth curves in these latter cases are very close to each other, 
as the effect of actuators on the oblique wave is
symmetric. In the antisymmetric streamwise vortex case, the most effective components are $v$ and $w$ (reduction by a factor $\approx6.3$ and $\approx9.9$,
respectively), whereas $u$ has a quite poor performance (amplification reduced by a factor $\approx1.2$). 
The differences in performance between $u$ and $w$ may be interpreted with a geometric argument, as for a streamwise-invariant perturbation
the $u$ component acts in a weakly controllable direction. It is also noteworthy that, in all these cases, 
the control action reduces the time interval after which the initial disturbance gets to its maximum amplification.

\begin{figure}[!htbp]
\begin{center}
\includegraphics[width=0.45\textwidth]{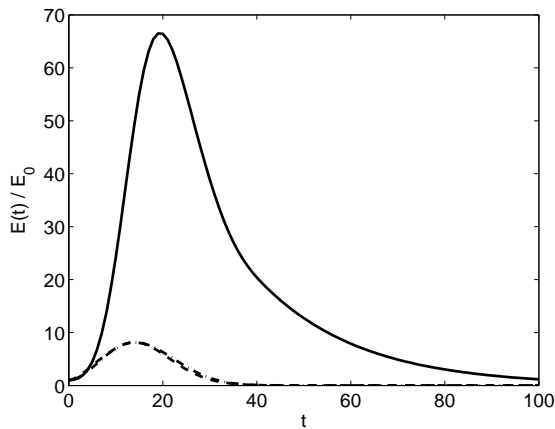}
\end{center}
\caption{Comparison of the linear evolution of the perturbation energy in unactuated case (--), closed-loop using LMI controller ($-\,\cdot$), 
and closed-loop using LQR controller ($-\, -$). Results for ($\alpha=1,\beta=1$) at $Re=2000$ ($\rho=0.5$), actuation with $v$, optimal
perturbation for the unactuated flow used as initial condition.}
\label{fig:lqr-ow}
\end{figure}

\begin{figure}[!htbp]
\begin{center}
\includegraphics[width=0.45\textwidth]{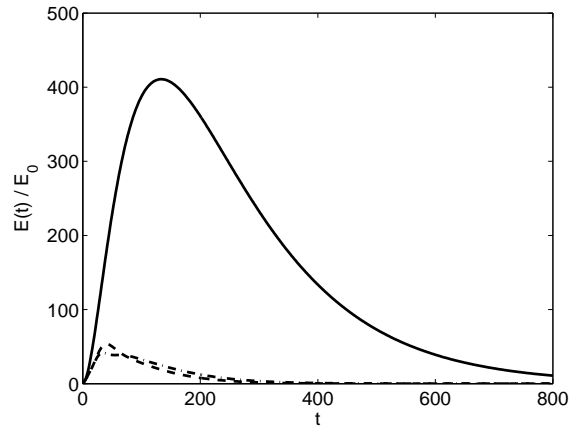}
\end{center}
\caption{Comparison of the linear evolution of the perturbation energy in unactuated case (--), closed-loop using LMI controller ($-\,\cdot$), 
and closed-loop using LQR controller ($-\, -$). Results for ($\alpha=0,\beta=2$) at $Re=2000$ ($\rho=8.0$), actuation with $w$, antisymmetric optimal
perturbation for the unactuated flow used as initial condition.}
\label{fig:lqr-sv}
\end{figure}

It is finally worth emphasizing that, when the closed-loop system experiences initial conditions in the form of optimal perturbations for the
unactuated flow, the LMI and LQR performance is practically equivalent, for a given global control effort. 
Considering the best performing LMI controllers ($v$ and $w$ actuation, respectively), results are given in fig. \ref{fig:lqr-ow} and \ref{fig:lqr-sv}, for the oblique wave and
streamwise vortex. 
For the optimal oblique wave (fig.\ref{fig:lqr-ow}), the
linear evolution of the perturbation energy using LQR control matches almost perfectly that obtained with the LMI controller. In the
case of antisymmetric streamwise vortex, we obtain a slightly larger maximum amplification for the LQR, that is however followed by 
a faster transient to zero if compared to the LMI case. The results reported in fig. \ref{fig:lqr-ow} and \ref{fig:lqr-sv} are substantially independent on further 
decrease of the value of $\rho$: no significant changes in the time evolution of the perturbation energy are obtained, but at a far larger expense.
This indicates that these results are close to the limit of small control weight.

\subsection{Closed-loop transition thresholds}

After these numerical experiments in the linear setting, the performance of LMI-based controllers
in terms of transition delay capabilities has been verified using Direct Numerical Simulations (DNS)
of transitional Poiseuille flow at $Re=2000$, using an existing computer code and computing
system \cite{luchini-quadrio-2006}. Controllers are tested against initial conditions
in the form of:
\begin{itemize}
\item a pair of oblique waves $(\alpha_0 = 1, \beta_0 = \pm 1)$, in a box of size $2\pi \times 2 \times 2\pi$;
\item antisymmetric streamwise vortices $(\alpha_0 = 0, \beta_0 = 2)$, in a box of size $2\pi \times 2 \times \pi$.
\end{itemize}
These initial conditions are obtained by computing the optimal
perturbations for the unactuated flow at the corresponding wavenumber pairs, and specifically for the streamwise vortices the optimal initial
condition having antisymmetric (with respect to the centerplane $y=0$) distribution of wall-normal velocity is considered, as 
it provides the optimal transition time in the nonlinear case\cite{reddy-etal-1998}. 
Random noise, in the form of a random combination of the first $30$ Stokes modes -- ordered by decreasing
real part of the corresponding eigenvalues -- is added on the wavenumber array $(0,\pm1,\pm2)\alpha_0$ and $(0,\pm1,\pm2)\beta_0$,
and the noise energy is chosen as $1\%$ of the total perturbation energy. 
The resulting optimal perturbations are identical to those reported in previous works\cite{reddy-etal-1998,Hogberg-2003-149}.

In order to reduce the computational problem of control design to an affordable size, LMI controllers are designed on the same array of wavenumber pairs where
random noise is introduced. Furthermore, the control effort tuning parameter is set at $\mu =  10$, a value which is derived from preliminary 
tests and previous work on the control of the linearized dynamics of streamwise vortices\cite{thesis}. The value of the parameter $\mu$ has been kept constant in wavenumber space;
however, it should be emphasized that this parameter could be a function of the wavenumber pair -- 
thus providing room for optimization of the control performance. 

The performance of LMI controllers is quantified by evaluating the closed-loop transition threshold\cite{reddy-etal-1998} 
of a given initial condition, for all the wall actuation components.
The mixed spatial discretization (fourth-order, compact finite differences in $y$ direction and Fourier expansion
in $x$ and $z$ directions) employs $64$ grid points in $y$ and $16\times64$ modes in $x$ and $z$. Time integration is
performed via the usual semi-implicit approach, where nonlinear terms are advanced explicitly using a low-storage Runge-Kutta algorithm
whereas linear diffusion terms are advanced implicitly via a Crank-Nicholson scheme. Each simulation was run over a time
window of $2000$ nondimensional time units, that proved sufficiently long to ensure that a laminar or a turbulent
state was reached after the initial transient growth of the perturbation energy. In order to obtain the thresholds
reported in table \ref{tab:thresholds}, a bisection algorithm was employed; this procedure requires a large number of simulations,
corresponding to approximately $3$ months of CPU time.

The open-loop transition thresholds reported in table \ref{tab:thresholds} agree with previous findings\cite{reddy-etal-1998}.
Results summarized in the table indicate that the LMI controller is able to increase the transitional
energy of the initial conditions considered. In particular, a synthetic performance measure is indicated in the 
table as improvement factor (I.F.), corresponding to the ratio between the threshold energy computed in the controlled
case over that corresponding to the unactuated flow. In general, for both the oblique waves and the streamwise vortices,
actuation with the wall-normal velocity $v$ outperforms actuation with the other components. This behavior
is expected, as forcing with wall-parallel components affects the flow by means of viscous diffusion only, whereas
forcing with $v$ introduces an additional non-zero momentum flux at the boundary. The improvement factors associated
with the oblique wave case are higher than those pertaining to the streamwise vortex case, when using $u$ and $v$ actuation.
For the $v$-component case, this is coherent with previous works\cite{Hogberg-2003-149}, and can be interpreted
physically by the argument that targeting oblique waves mitigates the subsequent development of streamwise vortices,
therefore reducing the strength of the associated streak instability. 
The $u$ component provides the overall worst performance in the streamwise vortex case, a result that
can be interpreted as a consequence of the particular geometrical configuration. In fact, with
respect to a streamwise-invariant spatial structure, $u$ actuation works in an approximately null direction,
whereas actuation with $w$ is better-suited, as shown by its improvement factor. A similar geometrical interpretation can be given for the almost
equal improvement factors obtained with $u$ and $w$, when an pair of oblique waves is given as initial condition.

\begin{table}
\begin{center}
\begin{tabular}{|c|c|c|c|c|}
\hline
 & \multicolumn{2}{|c|}{Oblique waves} & \multicolumn{2}{|c|}{Streamwise vortices} \\
 & Thres. & I.F. & Thres. & I.F.  \\
\hline
\hline
Open-loop & $2.39\cdot10^{-6}$ & & $6.47\cdot10^{-6}$ &  \\ 
\hline
$u$ - actuation & $9.89\cdot10^{-6}$ & $\approx4.14$ & $8.57\cdot10^{-6}$ & $\approx1.32$ \\
\hline
$v$ - actuation & $3.04\cdot10^{-5}$ & $\approx12.72$ & $4.86\cdot10^{-5}$ & $\approx7.51$\\
\hline 
$w$ - actuation & $9.77\cdot10^{-6}$ & $\approx4.09$ & $3.86\cdot10^{-5}$ & $\approx5.97$ \\
\hline
\hline
\end{tabular}
\end{center}
\caption{Open-loop and closed-loop transition thresholds, as measured by DNS. The values of transitional energy are
given with with an uncertainty of $\pm3\%$. The column labelled I.F. indicates the improvement factor 
in the closed loop with respect to the unactuated case with an initial condition having the same spatial structure.}
\label{tab:thresholds}
\end{table}

Results reported in table \ref{tab:thresholds} for the thresholds obtained using $v$ actuation are in qualitative agreement with previous work using the LQR controllers\cite{Hogberg-2003-149};
however, in quantitative terms the LQR approach outperforms the present LMI approach. In particular, the improvement factors reported with LQR for oblique waves and streamwise
vortices at $Re=2000$ are $I.F.=102$ and $I.F.=10$, respectively\cite{Hogberg-2003-149}; therefore, LQR controllers seem to perform substantially better than LMI controllers
in presence of oblique waves as initial conditions. It should be emphasized, however, that such comparison is not entirely appropriate. In fact, even if the same energy norm
is used to quantify the magnitude of velocity disturbances, control
laws are designed with different parameters constraining the control effort. As a matter of fact, the linear results reported in fig. \ref{fig:lin-ow} and \ref{fig:lin-sv} show
that the LMI controller performs similarly to the LQR controller for a value of the LQR control weight $\rho=0.5$ and $\rho=8.0$, respectively. These values are different from
the value of $\rho=0.01$, used uniformly in wavenumber space in H\"ogberg et al.\cite{Hogberg-2003-149}. A fair comparison between the two approaches is not possible in this case;
therefore, it is impossible to draw a conclusive statement about the effectiveness of feedback minimization of transient growth versus 
feedback minimization of the disturbance energy in transition delay.

As the ultimate goal of LMI controllers is that of preventing transition to turbulence, it can be important
to quantify the energy efficiency of these controllers in the nonlinear case. 
For instance, given a transitional initial condition, it
is possible to compare the energy expenditure of the controller to prevent transition with the additional energy to be  
introduced into the unactuated flow to compensate for the increase in friction, over the same time window
(i.e. the time necessary for the transient in the controlled flow to die out).
Referring to the best performing cases in table \ref{tab:thresholds}, we consider actuation with wall-normal
velocity $v$ and, in the two cases, initial conditions having energy $\approx 3\%$ below the corresponding
closed-loop threshold. Using the present nondimensionalization, a conservative estimate of the energy required for the control action in the
time interval $[0,T]$ can be given by\cite{bewley-moin-temam-2001}:
\begin{equation*}
E_c = \frac{1}{V}\int_0^T\int_{A_{u,l}} (|\frac{v^3}{2}| +|pv|) \, dA\,dt, \\
\end{equation*}
where $p$ is the fluctuating wall pressure and $A_{u,l}$ the upper and lower blowing/suction surfaces,
whereas the additional energy required to drive the unactuated flow against the increased viscous drag
on the same time interval is given by:
\begin{equation*}
E_\nu = \frac{1}{V}\int_0^T\int_{A_{u,l}} \frac{1}{Re} \Big(\frac{\partial U}{\partial y}-\frac{\partial U_{lam}}{\partial y}\Big)\, dA\,dt.
\end{equation*}
The ratio $E_c / E_\nu$ reads about $1.39\cdot10^{-3}$ and $1.68\cdot10^{-3}$, for the oblique waves and 
streamwise vortex case, respectively. Furthermore, linear tests using $v$ actuation have shown that $E_c$ can be
of the same order of magnitude of the actual reduction in maximum transient energy growth. Hence, these results indicate
that, even if the energy expenditure due to the control action is comparable to that experienced in the linear amplification
of the optimal disturbance, it is nevertheless negligible if 
compared to the potential energy saving due to transition prevention. 

\section{Conclusions}

The present work has considered the design of full-state feedback controllers specifically targeting the transient energy
growth mechanism in laminar channel flow. It has been shown that full transpiration at both walls and full-state knowledge
are not sufficient to ensure a monotonically stable closed-loop system via a linear feedback law. 
Further, an advanced control design technique -- based
on a LMI formulation -- has been employed to design feedback controllers that have been tested in both the linearized
setting and in nonlinear, transitional flows. 

Linear tests indicated that the LMI strategy allows to obtain a consistent reduction of the maximum open-loop transient growth. At a given
global control expenditure over a time window sufficiently long for the perturbations to decay to zero, the LMI-controlled closed-loop system
experiences a lower transient energy amplification than a LQR-controlled closed-loop flow. However, in presence of an optimal perturbation
for the unactuated flow, the performance of the two control strategies is practically equivalent. Results obtained in the linear setting further indicate that, 
in the case of perturbations in the form of oblique waves and streamwise vortices, the most effective actuation components are $v$ and $w$, respectively.

In the nonlinear case, it has been found that these controllers are capable of increasing the 
threshold energy for transition when initial conditions are given to the flow in the form of oblique waves or streamwise vortices;
the effectiveness of different actuation components has been addressed, indicating that wall blowing/suction is most effective
in providing a higher closed-loop threshold energy. 
Additionally, in transitional conditions, LMI controllers prove to be energy-effective, as the energy
required by the control action is negligible when compared to the energy saving due to avoiding transition.

\bibliographystyle{unsrt}

\end{document}